# Design and Implementation of a Blockchain-based Consent Management System

Nathaniel Aldred
*Ontario Tech University*
Oshawa, ON, Canada
Nathaniel.aldred@ontariotechu.net

Luke Baal
*Ontario Tech University*
Oshawa, ON, Canada
Luke.baal@ontariotechu.net

Graeham Broda
*Ontario Tech University*
Oshawa, ON, Canada
Graeham.broda@ontariotechu.net

Steven Trumble
*Ontario Tech University*
Oshawa, ON, Canada
Steven.trumble@ontariotechu.net

Qusay H. Mahmoud
*Ontario Tech University*
Oshawa, ON, Canada
Qusay.mahmoud@ontariotechu.ca

*Abstract*—A blockchain is a distributed ledger forming a distributed consensus on a history of transactions. It is the underlying technology for the Bitcoin cryptocurrency, but there are many applications beyond the financial sector. With blockchain's built-in security and removal of the need for third party trust, blockchain has started to see some use within contract applications among other things. In this paper, we present the design and implementation of a permissioned-based blockchain third party consent management system, whose policy can be decided by a government agency. We have constructed a proof of concept implementation using Hyperledger Fabric to provide a service that allows end-users to control and consent to who manages their private information. We believe our solution meets the guiding principles of EU General Data Protection Regulation (GDPR). While our performance and usability evaluation are limited, our solution design and its implementation meet the 7 foundational principles of privacy by design.

*Keywords*— Blockchain, consent management, user privacy, GDPR, Hyperledger, privacy by design.

## I. Introduction

With the introduction of the General Data Protection Regulation (GPDR), a new large set of restrictions and rules have been placed upon data usage and collection, with large penalties if a business fails to comply. The GDPR has greatly increased the scope of how all businesses that process EU citizens data are affected. Businesses must provide a clear description of what is involved in their consent agreements, as well as having an easy method of consent withdrawal. In consideration of a customer's data rights, there are several subject rights. In an attempt to summarize the rights described in the GDPR policy, we will list them and give a small description. The first right, *Notification of Breach*, a company must notify customers as soon as a data breach were to happen. *Right to Access* is the right that a customer must be notified if there data is being accessed for processing. The *Right to be Forgotten*, is the customers right to revoke access of a company to their data. The GDPR also promotes privacy by design which we will describe in more detail in the following paragraph. Finally, the GDPR will have Data Protection Officers, a group of people who will ensure that companies are complying with the appropriate data policies [1].

Even outside of the EU, there are many reasons why businesses would want to adopt these rules. In Canada, surveys have shown that it is a potential business advantage to leverage these rules to further protect private and secure data, such as the Office of the Privacy Commissioner of Canada [2]. Secure data can be defined in several ways. Any personal information relating to a customer is considered confidential and secure, so no one should be able to access it without permission. Information in this category includes, but is not limited to: photographs and IP addresses, contact lists, voice-print biometrics used in voice recognition apps, location information. This information can be collected by a malicious entity, combining many pieces of information creates an overview and insight into a victim's life. When rules are in place to protect this data, customers have been shown to feel significantly more confident in the company having their data, and their capabilities to protect it. Some examples of rules are in place to protect customers include application liability, which states that "Under Canada's private sector privacy legislation, an organization is accountable for personal information that it collects, uses and discloses", meaning that a business has a capital interest in treating customer information with care, and protecting it. All of this has significant impact on how software and data management systems are designed, since previously companies had no reason to get rid of data upon a user's request.

This now leads developers who would like to develop software using appropriate privacy policies, with several choices. The first choice is to develop their own in-house solution, developing their own privacy policy, and assuring customers that they are being upfront and honest in their dealings. This solution may not be adequate for many customers, as there is no way to ensure the company is handling their data as they say. The next option, would be to use a third party solution. This solution would need to have its policies decided by a government organization, and companies wishing to participate would need to be accredited by this organization. This would ensure that companies complied with the given policies, and that the policies could be made or altered by the elected party in accordance with the citizens wishes. This leads us to our next problem, the system enforcing these policies must be open and transparent about its privacy practices, as outlined by the Office of the Privacy

Commissioner of Canada. Companies and users participating in this system will want to ensure that the history of their transactions is immutable, yet remain private. Blockchain seems like an excellent way to implement several of these features. The blockchain is an immutable ledger, that provides a history of transactions, but allows the details to remain private [3]. It is a peer-to-peer network (in a permissionless blockchain) that removes the need of a third-party to authorize transactions. This fulfills most of our requirements but does not allow for a governing body to set policy on these transactions. In order to do this, we believe a permissioned blockchain with immediate finality [4] such as Hyperledger Fabric is needed.

In this paper, we present the design and implementation of a permissioned-based blockchain third party consent management system, whose policy can be decided by a government agency. We will mostly be discussing the technical implementation of such a system, unless its implementation is directly affected by government policy. In order to do so, the remaining of this paper is organized as follows. In Section II, we will present an overview of the background and related work. Section III introduces and discusses the design and implementation of our proposed solution. Evaluation results are presented in Section IV. Finally, conclusions and ideas for future work are presented in Section V.

## II. BACKGROUND AND RELATED WORK

One of the main premises that blockchain is built on, is the fact that security built into it removes the need for the trust of a third party. The trust is placed upon a distributed set of actors which as the added benefit of also making it hard to trace-back to a user from a permission record of theirs. This reduces the impact of trust since many of the actors have different interests making a malicious consent hard to coordinate. Also, the immutability makes a blockchain a great solution for consent management as once a permission is granted it is recorded in the ledger and becomes immutable, the way to revoke this permission is to add another record to the chain stating that it has been done so. It is also possible to keep transactions anonymous, despite them being announced publicly in a permissionless blockchain. The public keys can be kept private, and the public will only be able to see that there is a transaction, and will not know the who the user of the public key is [3].

Hyperledger is self coined, as a distributed ledger technology (DLT) [5]. It is an implementation of a modular blockchain architecture. It provides many features that we find may be suitable for developing a possible implementation of a consent management system. Hyperledger Fabric provides an identity management system, which is to say they require every user to be authenticated to participate and transact on the blockchain. This means that different levels of permissions can be applied to different users [5]. The extra level of permission management allows a centralized authority to allow certain users to transact with the blockchain in different ways. For example, a company may only be given read permission, and only users can manipulate the global state. This may at first seem to go against the idea of blockchain as a peer to peer network. However, if we look at the perspective of a government trying to regulate a privacy policy in order to protect their citizens rights, then a privacy policy that can be reinforced through code and developed by lawmakers, is an ideal solution. Additionally, due to the use of authentication (permissioned blockchain), compute intensive consensus algorithms such as the proof-of-work concepts are not needed, which makes Hyperledger a more scalable option.

Blockchain and its many applications are an active area of research. Most of the research related to privacy and consent management is being done in the medical industry, as patients' data is very lucrative in this field, and their rights must be protected. Genestier et al. [6] developed an application for a distributed consent management in using patient's information. In their model, the user can give permission for 3rd party companies to collect information on the patient, whether this is the use of records, or real time monitoring of the patient. Hyperledger was used for many of the same reasons we selected Hyperledger too. The second example of research being done is by Benchoufi et al. [7, 8] who have proposed using a blockchain solution in clinical trial research. Apparently, there has been some trouble with ensuring stakeholders do not manipulate results, and ensuring that trial properly collect consent from trial members, and properly inform the members of any new information and renewal of product given in testing. Finally Gammon [9] described the benefits of consent management using blockchain, and some similar technologies that already exist. There are several Medical related solutions [9]. In Longenesis [10], they use a decentralized system to provide a medical record marketplace. Selling your data is essentially exchanging value on the blockchain, which is what the blockchain excels at. Nebula Genomics [11], is under a similar branch, as they will purchase a user's entire genome, especially if the patient has unique health conditions. MedRec [12], which aligns with Genestier et al.'s design [6], and is used for sharing medical records. Estonian e-health has also already implemented a blockchain system giving patients access control of their data [9].

## III. PROPOSED SOLUTION

Before presenting the proposed solution, we will discuss the assumptions we have made and then present the architecture of the solution and the proof of concept prototype we have implemented, followed by the challenges we have faced.

### A. Assumptions

The major assumption made is that our proposed solution will be supported by a trusted organization. For example, ISO/PC 317 [13] Consumer protection: privacy by design for consumer goods and services, could provide an accreditation for a privacy policy management system. This accreditation would be similar to ISO 9001:20015 [14] which requires an organization to meet several quality requirements. Using this accreditation system and Hyperledger Fabric permissioned based system, we could require that companies wishing to participate in our blockchain are accredited. The policy would need to include several requirements from the companies participating. Companies would need to comply with any updates to user permissions. As an example, if a user were to revoke access their phone number the company would need to delete it from their internal database. Or even better yet, the company would only access the data in our database, updating the ledger in a request for a key, thereby alerting the user that their data is being accessed. Once a company has accessed a users data, they would be able to keep it forever, but should not mean that they can use it in any context they wish – only for the context the data was collected for. A development of a good privacy policy would entail the particulars of how that data is handled, and how to hold a company liable for it. This includes any breaches to the users data, and would motivate companies to leave the storage of data to the centralized database of the server. We are also under the assumption that users would not want their permission settings to be public to other companies. This means that given a permission asset, it must be impossible to determine what user that belongs to. The final assumption that we make, is that companies are not concerned with being connected to permission item, thereby reducing computational complexity.

### B. System Architecture

The overall architecture of our solution can be described as having three main components: The blockchain network, the REST API consortium (composer REST API, our REST API, and front-end), and the external database. Figure 1 depicts the current and implemented architecture of our system. As you can see, both the user and the companies need to register to utilize the system. The front-end handles all the requests to Hyperledger, and all requests for data from the back-end. These interactions will be described in more detail in the following sub-sections.

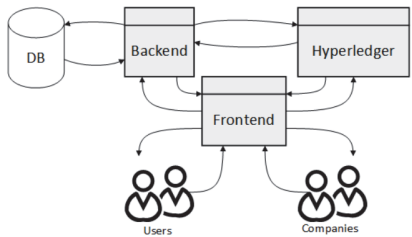

Figure 1: System architecture

We also considered an alternate architecture (Figure 2), where each user would have their own application to interact with Hyperledger and the data requests from the back-end. This would be to allow a company to interact with our application through the use of an API. Companies would be able to integrate our solution into their own applications how they would like, and have all interactions controlled programmatically.

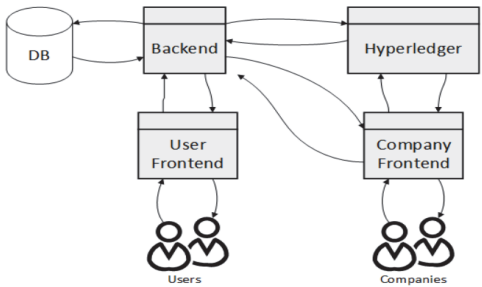

Figure 2: Alternative system architecture (not implemented)

#### 1) Blockchain Network

Hypderledger Fabric, which is the underlying blockchain is configured and managed by Hyperledger composer [5, 15]. Some configurations of Fabric are not done by composer and this is the network architecture. This is in regards to the organization, orderer, peer, and CouchDB modules. To handle this configuration, a basic sample network was used as it fills the needs of the prototype. The fabric network used is illustrated in Figure 3. It involves one organization, which is responsible for the management of the blockchain network, which is used by Composer. It also uses one orderer node, which is responsible for the creation of the blocks that get put on the blockchain, this is abstracted from the peer nodes in order to improve performance [15]. The network also features one peer node, which is what hosts the ledger or the blockchain. Finally, a CouchDB node is used as the data store in order to manage data and allow data persistence.

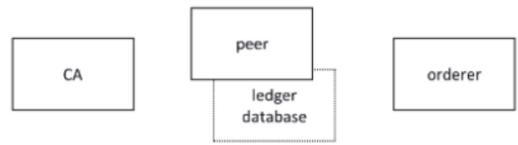

Figure 3: Hyperledger Fabric basic network

#### 2) REST API

In order to perform operations on the network or to just query data, Hyperledger Composer allows for the generation of a REST API to handle requests [15]. For simplicity, the REST API uses an Admin credential card with full access to the network to perform requests. The idea being to use an external server/application as a gateway to this REST API. Our own REST API was developed in order to make use of the database and to facilitate authentication for the front-end which was developed to allow users to register/login and add/edit permissions as well as view companies. Companies can also register/login to the front-end to view the data they have been given permission to see.

#### 3) External Database

Another key issue, as previously mentioned, is implementing GDPR's "right to be forgotten" within a blockchain system. The actual blockchain within Fabric is an immutable ledger, this means all versions of every asset are always going to be in the blockchain, with the most up-to-date versions being in the global state of Fabric [5]. Hence, storing the user's information in the blockchain is not viable, since once it was added, it could never be deleted, and might be visible to unauthorized parties. To solve this, a separate database is used to store the actual user information, which is why the permission assets in the blockchain use Boolean flags instead of the real information. A Mongo database was used as the database [16]. This database also generates a unique ID for each user entry, which is the user ID used to make pair key hashes described earlier.

### C. Implementation

The first question to be answered to implement this system was what should and should not be stored in the blockchain. As mentioned in our assumptions, users should not be able to be traced from their permissions, nor should companies be able to trace users they have to the permissions said users have set for other companies. It is still required, however, to be able to link users to their permissions entries, but only to authorized parties. Hyperledger Fabric allows for data to be represented as key-value pairs [5]. Assets are therefore required to be identified by a reference field, which will act as the key. For the permissions assets, the reference field is a hash of the user's ID and the company's ID. The assets also have a reference to the company they are associated with, which allows for the company's information to be retrieved upon querying the permissions, rather than just getting the company's ID and requiring a second query to get the name of the company. The remaining fields of the permissions assets are Boolean flags which indicate whether the company is allowed access to the respective information. For example, *"email: false, name: true"*, indicates the company is allowed access to the user's name, but not their email address. The other kind of data in the blockchain network is companies. It is assumed companies want their information to be readily available, so it was put into the blockchain, which better allows permissions to be linked to them. A company entry is simply some information fields designed to inform the user about the company, as well as a unique ID, which is implemented with UUID version 4. This unique ID is the company ID used in the pair key for the permissions assets.

The login information, including the users email or the companies name, as well as the password hash (plus a salt) of the user/company are stored in a database. Our REST API is then used to interact with this database. The REST API is responsible for handling user/company registration and login. Users enter their information on registration to have it added to the database. Companies, upon registering, only need to give their company name and a password. Once they have logged in after registering, they are prompted to enter their company information, which, along with their company name, is added to Hyperledger. Once a user logs in, they can view the list of companies in the system, so they can better identify the companies they want to give permissions to. They can then add permissions setting for a given company, as well as edit existing ones. Finally, users can also delete their account, which sets all permission settings to all be false, deletes their information from the database, and then logs the user out. It is worth noting the user will be warned and prompted to type "DELETE" and click confirm before any deletion occurs. The Hyperledger Composer REST API does allow for delete requests, however, we opted for setting all permissions to false, for two reasons. Firstly, the underlying blockchain will always have the history of the permission settings, so true deletion is not possible. Secondly, companies would have to recognize that permission entries were deleted in order to determine which set of information to delete. By setting all permissions to false, a message can be given to the companies telling them that the entry with the given pair key has been deleted, so they recognize that they must delete the information they have with that pair key. In order for companies to see the data they have access to, first the front-end sends a request to the REST API. The REST API then queries the Composer REST API for all permissions assets that have the same company ID as the currently authenticated company. The REST API uses the permission flags to determine what user information they get from the database and to send back to the front-end. Upon receiving the information, the front-end then renders each permission entry separately. The pair keys can be used by the companies to identify the individual entries.

### D. Scalability

The underlying blockchain software (Hyperledger) we used in our implementation allowed our system to be scalable. Hyperledger Fabric, which is the actual blockchain system, operates using several docker containers [5]. Having more peers improves the availability, given each peer hosts an instance of both the ledger and any chain code. This availability, however, is limited to querying. Any updates to the ledger require all peers to reach consensus to ensure the transaction is valid. Adding more peers, therefore, will slow down updates, but it also makes it harder for the network to be fooled by a malicious user and prevents the system from having a single point of failure [5].

The REST server generated by Composer can also be used as a Docker container. Being a REST API, the server can simply be duplicated and have multiple instances of itself running to handle more requests. A common way to do this efficiently would be to use either Docker Swarm [17] or Kubernetes [18]. Likewise, our REST API can be similarly scaled using Docker Swarm or Kubernetes. The front-end can also be served by our REST API, allowing it to have more availability too.

### E. Challenges

There were many challenges in developing and implementing the proposed solution, which mainly fell in to three main categories: overhead of learning concepts and technology related to blockchain, technical issues dealing with the various implementations of that technology, and ensuring permissions cannot be traced back to users. The solution to the first challenge was just due diligence on our part. We researched any problems or misunderstandings we had in order to fill out our understanding of the subject. For example, if all transactions are transparent, how will a user remain private? It turns out this can easily be done by keeping the public key anonymous [3]. We also had to research several development platforms and attempt a preliminary design in order to flush out possible problems with the platforms. For example, Ethereum, did not have the permission-based control we felt we needed. The next problem we ran into was understanding and developing on the platform we decided on, Hyperledger. We realised right away that developing on Hyperledger would be a very grueling and painstaking endeavour, if we built from the ground up. We found a solution to this by using Hyperledger Fabric, as a framework for building on. However, this still required vast amount of knowledge, and quite a large amount of work to build the appropriate network for our application, and that is without even taking our original objective into consideration. In order to circumvent this problem, and focus on our objective, we found Hyperledger-Composer, which uses the Hyperledger Fabric framework, but is easily configurable. Obviously, this caused its own set of problems. This is because such a high level of abstraction requires a large knowledge of the underlying components when errors and bugs start to occur during development. One instance of this was when we found that Hyperledger was not responding. When we traced the error through the extensive system, we found that it was possibly because of GRPC connection timeouts. Obviously a more in-depth knowledge of the system, built over time of development on it, would give us a better idea of what was going wrong, and how to troubleshoot it. The end result was that it there was a comma missing in the configuration file that relates to the GRPC connection. The third challenge we faced, was a set of problems that came with our actual objective. We needed to keep a reference to users in permission items, without having them traceable to the user. The solution to this was that we kept an id of permission items in a hash of the user id and the company id. In this case, the user id must be known to trace the user to the corresponding permission item, which can only be found in our database. The second issue in this set, was implementing the "right to be forgotten" to allow users to delete their account and data. This is a problem when it comes to blockchain and more specifically Hyperledger Fabric, given the ledger is an immutable history of transactions [5]. The solution we developed to accomplish this was to update all permission to false, thereby requiring a change in the global state, and then deleting the user from our database. This solves the problem, because once the user id is gone, there is no way to trace it back to the user.

## IV. EVALUATION RESULTS

The evaluation of the performance and usability is limited. The POST/PUT requests to composer's REST API seem very slow. The requests tend to take roughly 3 seconds, even when all nodes are on localhost. Hyperledger Fabric is a permission based blockchain, and is able to use more computationally efficient methods than proof-of-work to reach consensus, however, these methods still take a considerable amount of time. All transactions must reach a consensus by a pre-set number of nodes in order to change the global state.

For the usability evaluation, while we have not done any actual user experiments, our prototype implementation seems to fill many of the requirements set out by the Office of the Privacy Commissioner of Canada [2] in their design for privacy. The interface is simple, and the user can give permission just as easily as they can remove it. The user can browse companies in a single list, and add/edit permissions as they wish. The interface for companies is also very simple, and not much work is needed beyond writing a script to get whatever information they can. They can use our rest API to automate the process, or just use our front-end as their needs dictate. Overall, we feel this project is successful in building the groundwork for a third party regulated consent management system, and feel there is a great deal of applications this can be used for.

We also evaluated our proposed solution based on the Privacy by Design 7 Foundational Principles as outlined in [19]. The 7 foundational principles are: proactive not reactive: preventative not remedial; privacy as the default setting; privacy embedded into design; full functionality: positive-sum, not zero-sum; end-to-end security: full-lifecycle protection; visibility and transparency: keep it open; respect for user privacy: keep it user-centric.

### A. Proactive not reactive

In order for a company to be able to be a part of this system they must be accredited first by a set of criteria so that way we can validate their processes in terms of handling the data that is allowed to them through the system. This way we can assure privacy and security of the data before it is given out, instead of afterwards when they have already been given the data.

### B. Privacy as the deafult setting

Privacy as the default setting means a user's information is private by design and if they want to make it available, they need to explicitly allow it [19]. For example, for a social networking site, users cannot see your information unless you allow other users to see it, rather than the default being they can see it and you must explicitly turn it off. With this definition in mind, our system follows this principle, given that a company has no knowledge whatsoever about a user until they add permissions for that company. Even then, the company can only see the information they have been given access to by the user.

### C. Privacy embedded into design

Our system was built with privacy and security as top priorities. The system was designed from the ground up using privacy and security principles to ensure that privacy was embedded into our design. Each element within our architecture keeps these ideas in mind, and we ensured that no sections violated these principles.

### D. Full functionality

This principle is based off the assumption that there is typically a trade off between privacy and security [19]. In our system, we take this principle to heart, and have tried to create the highest amount of security and privacy possible. With our design, the data is only able to be accessed with user permission, and can only be understood with user consent, making our system meet both the requirements for security and privacy.

### E. End-to-end security

This principle involves the assurance that all aspects of the system are secure [19]. Our system, being a proof of concept prototype, does not meet this fully, given there is little security between the back-end and database, as well as between the front-end/back-end and the Composer REST API. This was done for convenience and the technologies involved all allow for much better methods of security, which could certainly be utilized. Therefore, even though our system as it stands does not meet complete end-to-end security, it would be very reasonable for it to meet this principle if needed.

### F. Visibility and transparency

The idea being visibility and transparency is to ensure a system is "secure" simply because no one knows, except the creators of the system, how it works [19]. By making the system open source for example, you allow others to verify the security of the system [19]. For our project to follow this principle we would need to make it open source. All the tools we used, specifically Hyperledger, are all open source tools, allows this system to follow this principle.

*G. Respect for user privacy*

The idea for this principle is that the user should be involved and have a significant say in their own privacy [19]. Our system design directly implements this idea by giving the user full control over their data and privacy. The user is the only one who can access their data without explicit permission, maintaining this principle.

## V. CONCLUSION AND FUTURE WORK

In this paper we have presented the design and implementation of a system that we believe meets the guiding principles set by GDPR regarding data privacy. Blockchain presents inherent issues in regard to the removal of data, however our system has provided a secure way to avoid putting blockchain in the first place, while maintaining the privileges of data access in the secure blockchain. Our system makes sure that privileges are kept secure and anonymous on the blockchain, with no way of retrieval after the customer removes their data. this provides a succinct way to protect customer rights, while maintaining the integrity of the blockchain. Our implementation source code is freely available and can be downloaded from [20].

In addition, we have identified two potential use cases where our solution would be able to be used in real world applications. First, our solution could be used in medical and clinical trials. Using our solution would ensure stakeholders do not tamper with trial results and ensure test subjects are informed on every step. This would enhance patient privacy and control over their data, while simultaneously guaranteeing the validity of the data being used in the experiments.

The second use case we have identified, is having a company which customers can use to store their data, which third part companies would have to go through to access their data. Instead of each company holding on to their individual copy of the data, the customer data could be consolidated in a single location, with the customers in control of permissions. This of course bring us back to the idea of a committee ensuring that any company using the database, is properly accredited by a standards organisation.

For future work, there are many areas where it can be scaled up and improved. The first area that we can work on, is the scaling of Hyperledger. By design, our solution should be able to scale to a fairly large size. While doing so, it must remain responsive and practical. This will need to be simulated and tested as best as possible. The next area is to provide access cards for a user/company to set their permissions with. This will make it easier to authenticate with the composer REST server. Another addition would be the use of a second key pair in transactions as an additional firewall. This would prevent the ability to link transactions to a single public key [3]. The final area we would like to see work done, is that of developing a standard privacy policy, and ensuring that any companies partaking in the blockchain consent management system are properly accredited.

## REFERENCES

[1] "Key changes with the general data protection regulation – eugdpr," https://eugdpr.org/the-regulation/, (Accessed on 04/09/2019).

[2] "Canadian businesses and privacy-related issues -of.ce of the privacy commissioner of canada," https://www.priv.gc.ca/en/ opc-actions-and-decisions/research/explore-privacy-research/2012/ por 2012 01/, (Accessed on 04/09/2019).

[3] S. Nakamoto et al., "Bitcoin: A peer-to-peer electronic cash system," 2008.

[4] C. Cachin, "Architecture of the hyperledger blockchain fabric," in Workshop on distributed cryptocurrencies and consensus ledgers, vol. 310, 2016.

[5] "Why hyperledger fabric? — fabricdocs 1.0 documentation," https://fabrictestdocs.readthedocs.io/en/latest/whyfabric.html, (Accessed on 04/09/2019).

[6] P. Genestier, S. Zouarhi, P. Limeux, D. Excof.er, A. Prola, S. Sandon, and J.-M. Temerson, "Blockchain for consent management in the ehealth environment: A nugget for privacy and security challenges," Journal of the International Society for Telemedicine and eHealth, vol. 5, pp. GKR–e24, 2017.

[7] M. Benchou., R. Porcher, and P. Ravaud, "Blockchain protocols in clinical trials: Transparency and traceability of consent," F1000Research, vol. 6, 2017.

[8] M. Benchou. and P. Ravaud, "Blockchain technology for improving clinical research quality," Trials, vol. 18, no. 1, p. 335, 2017.

[9] K. Gammon, "Experimenting with blockchain: Can one technology boost both data integrity and patients' pocketbooks?" 2018.

[10] "Longenesis," http://longenesis.com/, (Accessed on 04/09/2019).

[11] "How it works — nebula genomics," https://www.nebula.org/how it works, (Accessed on 04/09/2019).

[12] "Medrec," https://medrec.media.mit.edu/, (Accessed on 04/09/2019).

[13] "Iso/pc 317 -consumer protection: privacy by design for consumer goods and services," https://www.iso.org/committee/6935430.html, (Ac-cessed on 04/09/2019).

[14] "Iso 9001:2015 -quality management systems – requirements," https://www.iso.org/standard/62085.html, (Accessed on 04/09/2019).

[15] "Hyperledger composer," https://hyperledger.github.io/composer/latest/introduction/introduction, 2018, (Accessed on 04/09/2019).

[16] "Mongodb," https://www.mongodb.com/, 2019, (Accessed on 04/09/2019).

[17] Docker, "Swarm mode overview," https://docs.docker.com/engine/swarm/, 2019, (Accessed on 04/09/2019).

[18] T. L. Foundation, "Kubernetes," https://kubernetes.io/, 2019, (Accessed on 04/09/2019).

[19] A. Cavoukian, "Privacy by design: the 7 foundational principles", available online: https://www.iab.org/wp-content/IAB-uploads/2011/03/fred_carter.pdf.

[20] Source code for the implementation of the proposed solution: https://github.com/LukeBaal/CloudProject.